\documentclass[11 pt]{article}
\date{}
\author{\normalsize Ali~Tajer \normalsize \footnote{Electrical Engineering Department, Columbia University, New York, NY 10027  (email: tajer@ee.columbia.edu).}\and \normalsize Aria Nosratinia\footnote{Electrical Engineering Department, The University of Texas at Dallas, Richardson, TX 75083 (email: aria@utdallas.edu).}}

\usepackage{graphicx}
\usepackage{amsmath,amssymb}
\usepackage{cite}
\usepackage{url}

\newtheorem{theorem}{Theorem}
\newtheorem{lemma}{Lemma}
\newtheorem{corollary}{Corollary}
\newtheorem{remark}{Remark}

\def\dotgt{\;{\overset{\text{\Large .}}{\geq}}\;}
\def\dotlt{\;{\overset{\text{\Large .}}{\leq}}\;}
\def\doteq{{\overset{\text{\Large .}}{=}}}

\newcommand{\dff}{\stackrel{\scriptscriptstyle\triangle}{=}}
\newcommand{\med}{\;|\;}

\newcommand{\Heq}{\boldsymbol{H}_{\rm eq}}
\newcommand{\bx}{\boldsymbol{x}}
\newcommand{\by}{\boldsymbol{y}}
\newcommand{\bn}{\boldsymbol{n}}
\newcommand{\bg}{\boldsymbol{g}}
\newcommand{\bR}{\boldsymbol{R}}
\newcommand{\bW}{\boldsymbol{W}}

\newcommand{\bh}{\boldsymbol{h}}
\newcommand{\bH}{\boldsymbol{H}}
\newcommand{\bQ}{\boldsymbol{Q}}
\newcommand{\bI}{\boldsymbol{I}}

\newcommand{\bbf}{\boldsymbol{f}}
\newcommand{\balpha}{\boldsymbol{\alpha}}

\newcommand{\bLambda}{\boldsymbol{\Lambda}}
\newcommand{\bmu}{\boldsymbol{\mu}}

\newcommand{\bbe}{\mathbb{E}}

\newcommand{\dout}{{d_{\rm out}(R,\nu,L)}}
\newcommand{\dpep}{{d(R,\nu,L)}}

\newcommand{\snr}{{{\sf SNR}}}
\newcommand{\info}{{I}}
\newcommand{\pr}{P}
\newcommand{\pout}{P_{\rm out}(R,\nu,L)}
\newcommand{\pe}{P_{\rm err}(R,\nu,L)}

\begin{document}

\title{Diversity Order in ISI Channels with Single-Carrier
Frequency-Domain Equalizers\thanks{The material in this paper has been presented in part at the IEEE Globecom 2007.}}

\maketitle

\allowdisplaybreaks
\begin{abstract}
This paper analyzes the diversity gain achieved by single-carrier
frequency-domain equalizer (SC-FDE) in frequency selective channels, and uncovers the interplay between diversity gain $d$, channel memory length $\nu$, transmission block length $L$, and the spectral efficiency $R$. We specifically show that for the class of minimum means-square error (MMSE) SC-FDE receivers, for rates  $R\leq\log\frac{L}{\nu}$ full diversity of $d=\nu+1$ is achievable, while for higher rates the diversity is given by $d=\lfloor2^{-R}L\rfloor+1$. In other words, the achievable diversity gain depends not only on the channel memory length, but also on the desired spectral efficiency and the transmission block length. A similar analysis reveals that for zero forcing SC-FDE, the diversity order is always one irrespective of channel memory length and spectral efficiency. These results are supported by simulations.
\end{abstract}

\section{Introduction}
\label{sec:introduction}

\begin{figure*}[p]
   \centering
   \includegraphics[width=5.5in]{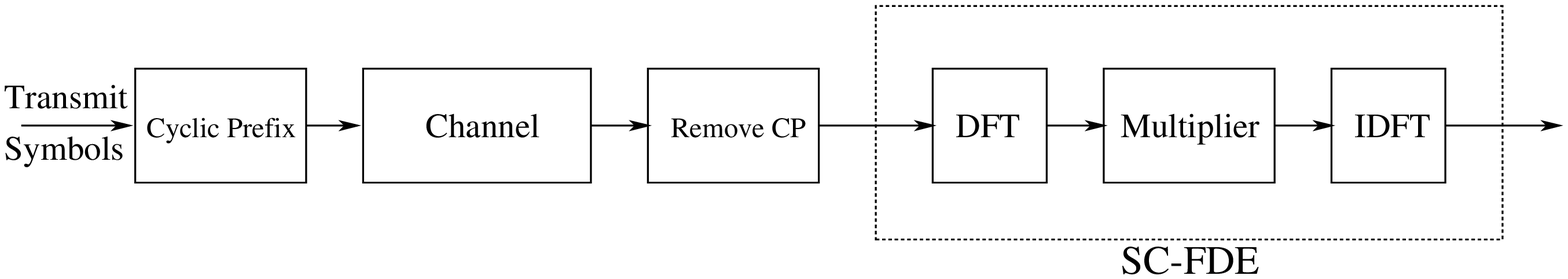}
   \caption{Block diagram of a SC-FDE system.}
   \label{fig:SC-FDE}
\end{figure*}

A single-carrier frequency-domain equalizer (SC-FDE), as depicted in
Fig.~\ref{fig:SC-FDE}, consists of simple single-carrier block
transmission with periodic cyclic-prefix insertion, and an equalizer
that performs discrete Fourier Transform (DFT) and single-tap filtering
followed by an inverse DFT (IDFT), where finally the equalizer output is fed into a slicer to make hard decisions on the input. Due to using
computationally efficient fast Fourier transform, SC-FDE has lower
complexity than time-domain equalizers.\footnote{This advantage is
especially pronounced in channels with long impulse response.}
Structurally, SC-FDE has similarities with OFDM, but has the key
distinction that SC-FDE decisions are made in the time domain, while
OFDM decisions are made in the frequency domain.
SC-FDE enjoys certain advantages over OFDM, as mentioned in,
e.g.,~\cite{Sari:MCOM95,Falconer:COMM02}. In particular SC-FDE is not
susceptible to the peak-to-average ratio (PAR) problem. Also, in OFDM
one must code across frequency bands to capture frequency diversity,
while in SC-FDE a similar issue does not exist since decisions are made
in the time domain. In addition, SC-FDE has reduced sensitivity to
carrier frequency errors, and confines the frequency-domain processing
to the receiver.
SC-FDE is deemed promising for broadband wireless
communication~\cite{Sari:MCOM95, Clark:JSAC98, Naofal:COML01,
Falconer:COMM02} and has been proposed for implementation in the 3GPP
long term evolution (LTE) standard.
This paper analyzes the SC-FDE and unveils hitherto unknown  relationships between its diversity, spectral efficiency, and  transmission block length. The explicit dependence of diversity on the transmission block length is especially intriguing, and to the best of our knowledge has no parallel in the literature of equalizers for dispersive channels.\footnote{In MIMO systems, a non-explicit dependence of diversity on block length is implied by the results of~\protect\cite{Zheng:IT03}}\footnote{Unlike \cite{Zheng:IT03} which uncovers the interplay between diversity and multiplexing gain (rates increase with $\log \snr$) we investigate the tradeoff between diversity and {\em fixed} rates. The results of \cite{Zheng:IT03} establish that for in MIMO flat-fading channels all fixed rates (corresponding to multiplexing gain 0) achieve essentially the same diversity. In contrast, we show that for ISI channels with SC-FDE changing the rate can affect the achievable diversity gain.}

We start by briefly reviewing some of the existing results on the
diversity gain of various block transmission schemes. It is known that
\emph{uncoded} OFDM is vulnerable to weak symbol detection when the
frequency selective channel has nulls on the DFT grid, and therefore, uncoded OFDM may not capture the full diversity of the inter-symbol interference (ISI) channel~\cite{Wang:IT03}. To mitigate this effect, various coded-OFDM schemes have been considered~\cite{Cioffi:COM92, Sadjadpour}. Motivated to achieve full diversity without error-control coding, \emph{complex-field coded (CFC)}-OFDM has been  introduced~\cite{Wang:IT03}, where it is shown to achieve full diversity with maximum likelihood (ML) detection. CFC-OFDM achieves its diversity in a manner essentially similar to the so-called signal space diversity of Boutros and Viterbo~\cite{Boutros:IT98}, by sending linear combinations of the uncoded symbols via each subcarrier. It has been shown that both zero-padded single-carrier block transmission and cyclic-prefix single-carrier block transmission are special cases of
CFC-OFDM~\cite{Wang:IT03}; therefore, by deploying {\em ML detection},
they also achieve full diversity.

The complexity of ML detection motivates the study of linear
equalizers. The first analysis on the diversity order of CFC-OFDM with
\emph{linear equalization} was provided in~\cite{cihan:IT04}, where it
is shown that with additional constraints on the code design,
zero-forcing (ZF) linear \emph{block} equalizers can achieve the same
diversity order as ML detection.  Furthermore, in~\cite{cihan:gc05} it
has been shown that zero-padded single-carrier block transmission, as a
special case of CFC-OFDM, meets the conditions discussed in~\cite{cihan:IT04} and therefore achieves full diversity by exploiting  ZF equalization.

Although it has been established that a cyclic-prefix single-carrier
block transmission {\em with ML detection,} achieves full
diversity~\cite{Wang:IT03}, the result clearly cannot be applied to
SC-FDE, because SC-FDE does not yield ML decisions. Furthermore the
linear equalization results mentioned in~\cite{cihan:IT04,cihan:gc05} do
not apply to SC-FDE either, since SC-FDE does not satisfy the conditions
in~\cite{cihan:IT04,cihan:gc05}. This distinction is further solidified
in the sequel where we show that SC-FDE in fact does \emph{not} enjoy
unconditional full diversity.

Our analyses reveal that for minimum-mean-square-error (MMSE) SC-FDE the
diversity order varies between 1 and channel length, $\nu+1$,
depending on the transmission setup. We demonstrate a tradeoff between
the achievable diversity order, data transmission rate, $R$
(bits/second/Hz), channel memory length, $\nu$, and transmission block
length, $L$. Specifically, at rates lower than $\log \frac{L}{\nu}$,
full diversity of $\nu+1$ is achieved, while at higher rates, the
diversity gain is $\lfloor2^{-R}L\rfloor+1$. These results support the
earlier analysis in~\cite{hedayat:asilomar04, Slock:ISIT06}, where it has been shown that for \emph{very low} and \emph{very high} data rates, diversity gains 1 and $\nu+1$ are achieved, respectively.  We also investigate the diversity order of zero-forcing (ZF) SC-FDE and find that the achievable diversity order is \emph{always} 1, which is similar to that of OFDM with zero-forcing equalization \cite{Zhang:SAM06}.

The rest of this paper is organized as follows. In
Section~\ref{sec:descriptions} the system model and some definitions
are provided. Diversity analysis for MMSE-SC-FDE and ZF-SC-FDE are
provided in  sections \ref{sec:diversity_MMSE} and
\ref{sec:diversity_ZF}, respectively. Section \ref{sec:simulations} provides numerical evaluations and simulation results and concluding remarks are presented in Section~\ref{sec:conclusion}.

\section{System Description}
\label{sec:descriptions}

\subsection{SC-FDE vs. OFDM}
\label{sec:SCFDE}

As seen in the baseband model of SC-FDE (Fig.~\ref{fig:SC-FDE}), after
removing the cyclic-prefix, a DFT operator is applied to the received
signal, each sample is multiplied by a complex coefficient and then an
IDFT transforms the signal back to the time domain. In the time domain,
the equalizer output is fed into a slicer to make hard decisions on the
transmitted vector.

In OFDM both channel equalization and detection are performed in the
frequency domain, whereas in SC-FDE, while channel equalization is done
in the frequency domain, receiver decisions are made in the time
domain, which leads to differences in the performance of
OFDM vs.\ SC-FDE. The underlying reason for such performance difference is that in uncoded OFDM, the subcarriers suffering from deep fade will exhibit poor performance. On the other hand, in SC-FDE detection decisions are made based on the (weighted) average performance of subcarriers, which is expected to be more robust to the fading of individual subcarriers. For more discussions see~\cite{Sari:MCOM95, Naofal:COML01}.

\subsection{Transmission Model}
\label{sec:model}
We consider a frequency selective quasi-static wireless fading channel
with memory length $\nu$,
\begin{equation*}
    H(z)=h_0+h_1 z^{-1}+\cdots+h_{\nu} z^{-\nu}.
\end{equation*}
The channel follows a block fading model where the channel coefficients
are independent complex Gaussian $\mathcal{CN}(0,1)$ random variables that remain unchanged over the transmission block of length $L$, and change to an independent state afterwards. Received signals are contaminated with zero-mean unit variance complex additive white Gaussian noise (AWGN). The channel output is given by
\begin{equation}\label{eq:model1}
    \by=\sqrt{\snr}\bH\bx+\bn,
\end{equation}
where $\bx=[x(L),\dots, x(-\nu+1)]^T$ denotes the transmitted block and
$\by=[y(L),\dots, y(1)]^T$ is the vector of received symbols before
equalization. We normalize $\bx$ such that the average transmit power
for each entry of $\bx$ is 1, and $\snr$ accounts for the average signal-to-noise ratio ($\snr$) at the transmitter. Channel noise is denoted by $\bn=[n(L),\dots, n(1)]^T$, and the channel matrix is represented by
\begin{equation}
 \label{eq:model2}
 \bH_{L\times(L+\nu)}\dff
\begin{bmatrix}
h_0 & h_1 & \cdots & h_{\nu} & 0 & \cdots & 0 \\ 0 & h_0 & h_1 &
 \cdots & h_{\nu} & \cdots & 0 \\ \vdots & \ddots & \ddots & \ddots &
 \ddots & \ddots & \vdots \\ 0 & \cdots & 0 & h_0 & h_1 & \cdots &
 h_{\nu}
\end{bmatrix}.
\end{equation}
To remove inter-block interference, a cyclic prefix is inserted at the
beginning of each transmit block, giving rise to the equivalent channel
\begin{equation*}
\bH_{\rm{eq}}\dff \begin{bmatrix}
  h_0 & h_1 & \cdots & h_{\nu-1} & h_{\nu} & 0 & \cdots & 0 \\ 0 &
  h_0 & h_1 & \cdots & h_{\nu-1} & h_{\nu} & \cdots & 0 \\ \vdots &
  \vdots & \vdots & \vdots & \vdots & \vdots & & \vdots \\ h_1 &
  h_2 & \cdots & h_{\nu}& 0 & 0 & \cdots & h_0
\end{bmatrix}.
\end{equation*}
This $L\times L$ circulant matrix $\bH_{\rm{eq}}$ has eigen
decomposition $\bH_{\rm{eq}}=\bQ^H\bLambda\bQ$, where $\bQ$ is the
discrete Fourier transform (DFT) matrix with elements
\begin{equation*}
 \bQ(m,n)=\frac{1}{\sqrt{L}}
 \exp\bigg[-j\frac{2\pi}{L}(m-1)(n-1)\bigg],\quad\mbox{for}\;\;m,n=1,\dots,L,
\end{equation*}
where we have $\bQ^H\bQ=\bI$. Also, the diagonal matrix $\bLambda$  contains the $L$-point (non-unitary) DFT of the first row of $\bH_{\rm{eq}}$ given by
\begin{equation}
 \label{eq:model3}
 \lambda_k\dff\bLambda_{k,k}=\sum_{i=0}^{\nu}h_i
 e^{-j\frac{2i\pi (k-1)}{L}},\quad\mbox{for}\;\;k=1,\dots, L.
\end{equation}
Each eigenvalue $\lambda_k$ is a linear combination of channel
coefficients, which are zero mean complex Gaussian random variables. Therefore $\{\lambda_k\}_{k=1}^L$ also have zero mean complex Gaussian
distribution.
\begin{remark} \label{remark:1}
For the special case of $L=\nu+1$, the eigenvalues
$\{\lambda_k\}_{k=1}^L$ are \emph{independent} random variables.
\end{remark}
We assume that the received signal is processed by a SC-FDE, designated by $\bW$, where its output $\tilde\by\dff[\tilde y(L),\dots,\tilde y(1)]$ is
\begin{equation*}
\tilde\by\dff \bW \by = \sqrt{\snr} \bW\Heq \bx + \bW \bn.
\end{equation*}

Throughout the paper we denote the transmission signal-to-noise ratio by $\snr$ and we say that the two functions $f(\snr)$ and $g(\snr$) are \emph{exponentially equal}, denoted by $f(\snr)\doteq g(\snr)$, when
\begin{equation*}
 \lim_{\snr\rightarrow\infty}\frac{\log f(\snr)}{\log\snr}=
 \lim_{\snr\rightarrow\infty}\frac{\log g(\snr)}{\log\snr}.
\end{equation*}
The ordering operators $\dotlt$ and $\dotgt$ are also defined  accordingly. If $f(\snr)\doteq\snr^d$, we say that $d$ is the \emph{exponential order} of $f(\snr)$.

\subsection{Diversity Analysis}
The diversity gain describes how fast the average pairwise error probability decays as the $\snr$ increases. For an ISI channel with memory length $\nu$ and SC-FDE receiver with block length $L$, we denote the diversity gain at data rate $R$ by $\dpep$ and is given by
\begin{equation}
\dpep \dff
-\lim_{\snr\rightarrow\infty}\frac{\log\pe}{\log\snr},
\label{eq:dpep}
\end{equation}
where $\pe$ denotes the average pairwise error probability, which is the probability that the receiver decides erroneously in favor of $s_k$, while $s_j$ was transmitted, i.e.,
\begin{eqnarray*}
\pe \dff {\bbe}\bigg\{P\big[s_j \rightarrow s_k\;| \bH=H\big]\bigg\}= {\bbe}\bigg\{P\Big[\|\tilde y(\ell)-\sqrt{\snr}s_j\|>\|\tilde y(\ell) -\sqrt{\snr}s_k\;\big| \bH=H\Big]\bigg\}.
\end{eqnarray*}

In this paper we aim to characterize $\dpep$, whose direct analysis requires a PEP analysis that depends on the choice of signaling. This approach is not easily tractable and as a remedy, we first turn to mutual information and outage analysis and characterize the exponential order of the outage probability. In the next step, by establishing that the outage probability and the average PEP exhibit identical exponential orders, we can characterize $\dpep$.

Therefore, we will also perform outage analysis for SC-FDE, whose related definitions are as follows. Due to the equalizer structure, the effective mutual information between $\bx$ and $\tilde\by$ is equal to the sum of the mutual information of their components (sub-streams)~\cite{onggosanusi:ICASSP02}
\begin{equation}
  \info (\bx;\tilde \by)=\frac{1}{L}\sum_{\ell=1}^L \info (x_\ell;\tilde y_\ell).
\label{eq:I}
\end{equation}
Subsequently, we define the following outage-type quantities
\begin{eqnarray}
\nonumber \pout & \dff & P\big[\info (\bx;\tilde\by)<R\big],\\
\mbox{and}\quad \dout & \dff &
-\lim_{\snr\rightarrow\infty}\frac{\log\pout}{\log\snr}.
\label{eq:dout}
\end{eqnarray}

\section{Diversity Analysis of MMSE SC-FDE}
\label{sec:diversity_MMSE}

We start with finding the unbiased decision-point ${\sf SINR}$. For the
transmission model given in~(\ref{eq:model1}) the MMSE linear equalizer
is
\begin{equation}
\bW_{\rm MMSE} =\big[\Heq^H\bH_{\rm
 eq}+\snr^{-1}\bI\big]^{-1}\Heq^H,
\label{eq:MMSE_filter}
\end{equation}
and the output of the equalizer can be found as
\begin{equation*}
 \tilde\by=\big[\Heq^H \Heq+
 \snr^{-1}\bI\big]^{-1}\Heq^H
 \Heq\bx+ \big[\Heq^H
 \Heq+\snr^{-1}\bI\big]^{-1}\Heq^H\bn.
\end{equation*}
We also define the noise term $\tilde\bn\dff[\tilde n(L),\dots,\tilde n(1)]$ as
\begin{equation}\label{eq:MMSE_noise}
  \tilde \bn\dff \tilde \by-\sqrt{\snr}\bx=\sqrt{\snr}(\bW\bH_{\rm eq}-\bI)\bx+\bW \bn,
\end{equation}
which accounts for the combined effect of the channel noise $\bn$ and the ISI residual due to MMSE interference suppression.
By recalling the eigen decomposition of $\Heq$ and noting that $\bbe[\bn]=\boldsymbol{0},\; \bbe[\bn\bn^H]=\bI$, some simple manipulations provide that
\begin{eqnarray}
  \label{eq:MMSE_mean1}\bmu_{\tilde\bn} & \dff &\bbe[\tilde\bn]=\sqrt{\snr}(\bW\bH_{\rm eq}-\bI)\bx,\\
  \label{eq:MMSE_var} \mbox{and}\quad\qquad \bR_{\tilde\bn} & \dff&\bbe[\tilde\bn\tilde\bn^H] = \big[\Heq^H  \Heq+\snr^{-1}\bI\big]^{-1}.
\end{eqnarray}
Due to the underlying symmetry, it can be show that the diagonal elements of $\bR_{\tilde\bn}$ are identical. Therefore, the unbiased decision-point ${\sf SINR}$ of MMSE SC-FDE for detecting symbol $x(\ell), 1\leq \ell\leq L$ (or the $\ell^{th}$ information stream) is
\begin{eqnarray}\label{eq:MMSE_snr}
 \nonumber  \gamma_\ell^{\rm MMSE}&\dff&\frac{\snr}{\bR_{\tilde\bn}(\ell,\ell)} -1 = \frac{\snr}{\frac{1}{L}{\rm tr}[\bR_{\tilde\bn}]} -1 = \frac{\snr}{\frac{1}{L}{\rm tr}  \big[\Heq^H\Heq+\snr^{-1}\bI\big]^{-1}}-1\\
 &=& \frac{1}{\frac{1}{L}{\rm
 tr}\big[\snr\;\bLambda\bLambda^H+\bI\big]^{-1}}-1= \bigg[\frac{1}{L}\sum_{k=1}^L\frac{1}{1+\snr|\lambda_k|^2}\bigg]^{-1}-1,
\end{eqnarray}
which does not depend on $\ell$ and is identical for all information streams. Therefore, the mutual information in (\ref{eq:I}) becomes
\begin{equation}
 \info_{\rm  MMSE}(\bx;\tilde\by) =\frac{1}{L}\sum_{\ell=1}^L \log(1+\gamma_\ell^{\rm MMSE})= -\log\bigg[\frac{1}{L}\sum_{\ell=1}^L\frac{1}{\snr|\lambda_k|^2+1}\bigg],
\label{eq:mmse_mutual}
\end{equation}
and the outage probability for the target rate $R$, which is the probability that the mutual information $\info(\bx;\tilde\by)$ falls below $R$ is
\begin{equation}\label{eq:Pout}
    \pout =\pr \bigg[\frac{1}{L}
 \sum_{k=1}^{L} \frac{1}{1+\snr|\lambda_k|^2}>2^{-R}\bigg].
\end{equation}

\subsection{Outage Analysis}
\label{sec:outage}

For analyzing the outage probability, we start with the special case of $L=\nu+1$, and then generalize the result for the arbitrary choices of $L$. The following lemma has a key role in finding the exponential order of the outage probability.
\begin{lemma}
\label{lemma:1}
For $n$ i.i.d. normal complex Gaussian random
variables $\{\lambda_k\}_{k=1}^n$ and a real-valued constant
$m\in(0,n)$ we have
\begin{equation}
 \label{eq:lemma1}
 \pr\bigg[\sum_{k=1}^n
 \frac{1}{1+\snr|\lambda_k|^2}>m\bigg] \doteq\snr^{-(\lfloor
 m\rfloor+1)},
\end{equation}
where $\lfloor \cdot \rfloor$ denotes the floor function.
\end{lemma}
\begin{proof}
We define
\begin{equation}
 \label{eq:alpha}
\alpha_k \dff -\frac{\log|\lambda_k|^2}{\log\snr},\quad\mbox{for}\;\;
 k=1,\dots,n \;\; ,
\end{equation}
based on which we can write the equality-in-the-limit
\begin{equation*}
 \frac{1}{1+\snr|\lambda_k|^2} \; \doteq \begin{cases}
 \snr^{\alpha_k-1} &  \alpha_k<1 \\
 1& \alpha_k>1
 \end{cases} \;\; .
\end{equation*}
This indicates that the term $ \frac{1}{1+\snr|\lambda_k|^2}$ is either 0 or 1 corresponding to the regions $\alpha_k<1$ and $\alpha_k>1$, respectively. Therefore, the probability in (\ref{eq:lemma1}) is exponentially equal to having at least $(\lfloor m\rfloor+1)$ number of $\{\alpha_k\}$ greater 1. In other words,
\begin{eqnarray}
\sum_{k=1}^n
 \frac{1}{1+\snr|\lambda_k|^2} &\doteq& \sum_{\alpha_k>1} 1 +
 \sum_{\alpha_k<1}\snr^{\alpha_k-1} \doteq M(\balpha)+\max_{\{\alpha_k\med \alpha_k<1\}}\snr^{\alpha_k-1},
\end{eqnarray}
where we have defined $\balpha\dff[\alpha_1,\dots,\alpha_{n}]$ and a new random variable
\begin{equation}\label{eq:M}
    M(\balpha) \dff \sum_{\alpha_k>1} 1 \; ,
\end{equation}
i.e., $M(\balpha) $ counts the number of $\alpha_k>1$. Clearly $\{\alpha_1,\ldots,\alpha_n\}$ and $M(\balpha) $
are random variables induced by $\{\lambda_1,\ldots,\lambda_n\}$. Knowing that $|\lambda_k|^2$ has exponential distribution, by using arguments
similar to~\cite{Zheng:IT03} it can be verified that the cumulative density function (CDF) of $\alpha_k$ is
\begin{equation}
\label{eq:alpha_CDF}
    F_{\alpha_k}(\alpha)\doteq\exp\big(-\snr^{-\alpha}\big).
\end{equation}
As a result $P(\alpha_k>1)\doteq1-\exp(-\snr^{-1})\doteq\snr^{-1}$. Invoking the independence of $\{\lambda_k\}$, and thereof the independence of $\{\alpha_k\}$, provides that the random variable $M(\balpha)$ is binomially distributed and its binomial parameter is asymptotically $\snr^{-1}$. Hence,
\begin{eqnarray*}
   \pr\bigg[\sum_{k=1}^n
   \frac{1}{1+\snr|\lambda_k|^2}>m\bigg] &\doteq& P\big(M(\balpha) +
   \max_{\{\alpha_k\med \alpha_k<1\}} \snr^{\alpha_k-1}>m\big) \\
   &\doteq& P(M(\balpha)>m)\\
   &=& \sum_{i=\lfloor m\rfloor+1}^nP\big(M(\balpha)=i\big)\\
   &\doteq&\sum_{i=\lfloor m\rfloor+1}^n{n\choose i}\snr^{-i}\underset{\doteq
   1}{\underbrace{\big(1-\snr^{-1}\big)}}^{n-i}\\
   &\doteq&\snr^{-(\lfloor m\rfloor+1)},
\end{eqnarray*}
In the above equations, the first (asymptotic) equality follows from
exchange of limit and probability due to continuity of functions, the
second equality holds because $\max_{\{\alpha_k\med \alpha_k<1\}}\snr^{\alpha_k-1}$ diminishes at high $\snr$, and the final equality follows from the fact that inside the summation the term with the largest exponent dominates. This concludes the proof of the lemma.
\end{proof}

Now, by using the above lemma, we offer the following theorem which
establishes the exponential order of the outage probability for
$L=\nu+1$.
\begin{theorem}[Outage Probability for $L=\nu+1$]
\label{th:1} In an ISI channel with memory length $\nu$, transmission block length $L=\nu+1$, data rate $R$, and an MMSE SC-FDE receiver, the outage probability satisfies
\begin{equation*}
 P_{\rm out}(R,\nu,\nu+1)\doteq \snr^{-d_{\rm out}(R,\nu,\nu+1)},
\end{equation*}
where
\begin{equation}
 \label{eq:th:mmse1}
 d_{\rm out}(R,\nu,\nu+1)=\lfloor 2^{-R}(\nu+1)\rfloor+1.
\end{equation}
\end{theorem}
\begin{proof}
Given the mutual information in~(\ref{eq:mmse_mutual}), for the case
of $L=\nu+1$ the outage probability is
\begin{eqnarray}\label{eq:th1_out1}
 P_{\rm out}(R,\nu,\nu+1) &=& \pr\big[\info_{\rm MMSE}(\bx;\tilde
 \by)<R\big]=\pr\bigg[\sum_{k=1}^{\nu+1}\frac{1}{1+\snr|\lambda_k|^2}>
 2^{-R}(\nu+1)\bigg].
\end{eqnarray}
As mentioned earlier in Remark~\ref{remark:1} for $L=\nu+1$, $\{\lambda_k\}$ are i.i.d. with complex Gaussian distribution. By setting $n=\nu+1$ and $m\dff 2^{-R}(\nu+1)$ for nonzero rates $R>0$, it is seen that $m\in(0,\nu+1)$. Therefore, the necessary conditions of
Lemma~\ref{lemma:1} are satisfied and consequently we have
\begin{equation}
 \label{eq:th1_out2}
 P_{\rm out}(R,\nu,\nu+1)\doteq\snr^{-(\lfloor
 2^{-R}(\nu+1)\rfloor+1)},
\end{equation}
which concludes the proof.
\end{proof}

In the next step, we generalize the results above for the arbitrary choice of block length $L$. We offer the following lemma which facilitates the transition from the special case of $L=\nu+1$ to any the arbitrary value for $L$.

\begin{lemma}
\label{lemma:2}
Consider the vector of channel coefficients $\bh\dff [h_0, \dots,
h_{\nu}]$ together with its two zero-padded versions $\bg_{1\times L}$ and $\bg'_{1\times L'}$
that differ only in the number of zeros padded, i.e.,
\begin{equation*}
\bg_{1\times L} \dff [h_0,\dots,h_{\nu},\underset{L-\nu-1}{\underbrace{0,\dots,
   0}}], \qquad\mbox{and}\qquad \bg'_{1\times L'}\dff[h_0, \dots, h_{\nu},
   \underset{L'-\nu-1}{\underbrace{0, \dots, 0}}].
\end{equation*}
The DFT vectors $\{\lambda_i\}_{1\times L}\dff{\rm
DFT}(\bg)$ and $\{\lambda'_i\}_{1\times L'}\dff{\rm DFT}(\bg')$
have the following property for any real-valued constant $m\in(0,\nu+1)$
\begin{equation}
 \label{eq:lemma2}
 \pr\bigg[\sum_{k=1}^L\frac{1}{1+\snr|\lambda_k|^2}>m\bigg]
 \doteq
 \pr\bigg[\sum_{k=1}^{L'}\frac{1}{1+\snr|\lambda'_k|^2}>m\bigg].
\end{equation}
\end{lemma}
\begin{proof}
See Appendix \ref{app:lemma3}.
\end{proof}

By using the above lemma, we generalize $d_{\rm out}(R,\nu,\nu+1)$ as characterized in Theorem \ref{th:1} for the general case of arbitrary $L$ to obtain $\dout$. The result offered in the next theorem demonstrates how altering the transmission block length from $L$ to $L'$ influences the characterization of $\dout$.

\begin{theorem}
\label{th:2} In an ISI channel with memory length $\nu$ and MMSE SC-FDE receiver, the exponential order of the outage probability of block transmission length $L$ and rate $R$ is equivalent to that of block length $L'$ and rate $R+\log\frac{L'}{L}$, i.e.,
\begin{equation}
 \label{eq:th2}
 \dout=d_{\rm out}\big(R+\log\frac{L'}{L},\nu,L'\big).
\end{equation}
\end{theorem}

\begin{proof}
By defining $\beta=\log\frac{L'}{L}$ we have
\begin{eqnarray}
 \label{eq:th2_proof1}
\pout &=&\pr \bigg[\frac{1}{L}
 \sum_{k=1}^{L} \frac{1}{1+\snr|\lambda_k|^2}>2^{-R}\bigg]\\
 \nonumber &=&\pr\bigg[\sum_{k=1}^{L}
 \frac{1}{1+\snr|\lambda_k|^2}>L2^{-R}\bigg]\\
 \label{eq:th2_proof2}
 &\doteq &\pr\bigg[\sum_{k=1}^{L'}
 \frac{1}{1+\snr|\lambda'_k|^2}> L2^{-R}\bigg]\\
 \nonumber
 &= &\pr\bigg[\frac{1}{L'}\sum_{k=1}^{L'}
 \frac{1}{1+\snr|\lambda'_k|^2}>\frac{L}{L'}2^{-R}\bigg]\\
 \nonumber
 &=& \pr\bigg[\frac{1}{L'}\sum_{k=1}^{L'}
 \frac{1}{1+\snr|\lambda'_k|^2}>2^{-(\beta+R)}\bigg]\\
 \label{eq:th2_proof3}
&=& P_{\rm out}\big(R+\beta,\nu,L'\big),
\end{eqnarray}
where~(\ref{eq:th2_proof2}) holds according to Lemma~(\ref{lemma:2})
for $m=L2^{-R}$. Exponential equality of
~(\ref{eq:th2_proof1})~and~(\ref{eq:th2_proof3}) shows that
\begin{equation*}
\snr^{-d_{\rm out}(R,\nu,L)}\doteq\snr^{-d_{\rm
out}(R+\log\frac{L'}{L},\nu,L')},
\end{equation*}
which completes the proof.
\end{proof}

Combining  Theorem~\ref{th:1} and
Theorem~\ref{th:2} leads to the main result of this paper as stated in the following corollary.

\begin{corollary}[Outage Probability]
\label{cor1} In an ISI channel with memory length $\nu$, transmission block length $L$, data rate $R$, and an MMSE SC-FDE receiver, the outage probability is characterized by
\begin{equation*}
 \pout\doteq\snr^{-\dout},
\end{equation*}
where,
\begin{equation}
 \label{eq:cor1}
 \dout=\begin{cases}
 \nu+1 & \mbox{\rm for}\; R\leq\log\frac{L}{\nu} \\
 \lfloor 2^{-R}L\rfloor+1& \mbox{\rm for}\;R > \log\frac{L}{\nu}
\end{cases}
.
\end{equation}
\end{corollary}
\begin{proof}
We use the result of the case $L=\nu+1$ as the benchmark. For this case
as given in (\ref{eq:th:mmse1}) we observe that for the rate interval
$(\log\frac{\nu+1}{i},\log\frac{\nu+1}{i-1}]$, we have $d_{\rm out}=i$, for $i=1,\dots,\nu+1$. By invoking the result of Theorem~\ref{th:2} and
setting $L'=\nu+1$ it is concluded that for block transmission length
$L$, the rate interval for which $\dout=1$ shifts to the interval
$(0,\log\frac{L}{\nu}]$ and the rate interval for which $\dout=i\geq 2$ shifts to the interval $(\log\frac{\nu+1}{i}+\log\frac{L}{\nu+1},\log\frac{\nu+1}{i-1}+\log\frac{L}{\nu+1}]=
(\log\frac{L}{i},\log\frac{L}{i-1}]$ for $i=2,\dots,\nu+1$. Such intervals can be mathematically represented as in (\ref{eq:cor1}).
\end{proof}

The analyses above convey that the maximum value of $\dout$ is
$\nu+1$ and is achievable for all rates not exceeding
$\log\frac{L}{\nu}$. If transmission rate increases beyond this point,
$\dout$ degrades following the rule given in (\ref{eq:cor1}). Such degradation can be compensated by increasing $L$.

\subsection{PEP Analysis}
\label{sec:PEP}

In this section, we find lower and upper bounds on $\dpep$ and show that
these bounds meet and are equal to $\dout$. The result is established
via two lemmas. We start by a lemma that utilizes the techniques
developed in~\cite[Lemma 5]{Zheng:IT03}. This lemma differs with \cite[Lemma 5]{Zheng:IT03} in the sense that we are dealing with rate, whereas \cite[Lemma 5]{Zheng:IT03} deals with multiplexing gain, and also the analysis of \cite[Lemma 5]{Zheng:IT03} exploits the fact that outage probability is continuous with respect to multiplexing gain, while in our analysis, as shown in Corollary \ref{cor1}, the outage probability is only {\em left}-continuous with respect to rate.
\begin{lemma}[Upper bound]\label{lemma:3}
For an ISI channel with MMSE SC-FDE receiver we have
\begin{equation*}
    \dout\geq\dpep
\end{equation*}
if $\exists\; d_{\min},\;d_{\max}\in\mathbb{R}_{++}$ such that $d_{\min}\leq \dpep\leq d_{\max}$.
\end{lemma}
\begin{proof}
We fix a codebook $\mathcal{C}$ of size $2^{Rl}$, where $R$ and $l$ are data rate and code length, respectively and $\boldsymbol x\in\mathcal{C}$ is the input to the system. The system input and output are related through the mapping $\tilde\by=\bbf(\bx)+\tilde\bn$, where $\bbf$ accounts for the combined effect channel and equalizer. All transmit messages are assumed to be equiprobable which provides ${\cal H}(\bx)=\log|{\cal C}|=Rl$, where ${\cal H}(\cdot)$ denotes entropy. By defining $E$ as the error event from Fano's inequality we get \cite[2.130]{Cover:book}
\begin{equation*}
{\cal H}(P(E)\med \bbf=f)+Rl\times P(E\med \bbf=f)\geq {\cal H}(\bx\med\tilde\by,\;\bbf=f).
\end{equation*}
Therefore,
\begin{equation}\label{eq:fano}
\pr({E}\med \bbf=f)\geq \frac{Rl- \info (\bx;\tilde\by\med \bbf=f)}{Rl}-\frac{\mathcal{H}(P(E)\med\bbf=f)}{Rl}.
\end{equation}
By defining $D_\delta$ for any value of $\delta>0$ as
\begin{equation*}
 \mathcal{D}_{\delta}\dff\{f: \info (\bx;\tilde \by\med\bbf=f)<l(R-\delta)\},
\end{equation*}
and noting that $\mathcal{H}(P(E)\med f\in{\cal D}_\delta)\leq \mathcal{H}(P(E))$ from (\ref{eq:fano}) we get
\begin{equation}\label{eq:fano2}
\pr({E}\med f\in{\cal D}_\delta)\geq \frac{Rl- \info (\bx;\tilde\by\med f\in{\cal  D}_\delta)}{Rl}-\frac{\mathcal{H}(P(E))}{Rl}\geq \frac{\delta}{R}-\frac{\mathcal{H}(P(E))}{Rl}.
\end{equation}
Also by using the definition of $\pout$ we have
\begin{equation}
\label{eq:lower1}
 \pr[f\in\mathcal{D}_\delta]=\pr\big[ \info (\bx;\tilde
   \by)<l(R-\delta)\big]\;\doteq\;\snr^{-d_{\rm out}(R-\delta,\nu,L)}.
\end{equation}
In our system (MMSE SC-FDE), we saw that function $\dout$ is  left-continuous with respect to $R$  since the ranges over which the diversity gains are constant are $(0,\log\frac{L}{\nu}],\dots,(\log L,\infty]$. Therefore, for small enough values of $\delta>0$, we have $d_{\rm out}(R,\nu,L)=d_{\rm out}(R-\delta,\nu,L)$. Hence, for for small enough values of $\delta>0$ by invoking (\ref{eq:fano2}) and (\ref{eq:lower1}) we have
\begin{eqnarray}\label{eq:UB}
 \nonumber \pe &=&\pr( E\med f\in\mathcal{D}_{\delta})\;\pr(f\in\mathcal{D}_{\delta})+
 \pr( E\med f\notin\mathcal{D}_{\delta})\;\pr(f\notin{\mathcal{D}}_{\delta})\\
 \nonumber &\geq& \pr( E\med f\in\mathcal{D}_{\delta})\;\pr(f\in\mathcal{D}_{\delta})\\
 \nonumber &\dotgt& \left(\frac{\delta}{R}-\frac{\mathcal{H}(P(E))}{Rl}\right)\; \snr^{-d_{\rm out}(R-\delta,\nu,L)}\\
 &\doteq&\left(\frac{\delta}{R}-\frac{\mathcal{H}(P(E))}{Rl}\right)\; \snr^{-d_{\rm out}(R,\nu,L)}.
\end{eqnarray}
Next we show that $\left(\frac{\delta}{R}-\frac{\mathcal{H}(P(E))}{Rl}\right)\doteq 1$. By recalling the definition of the diversity gain given in (\ref{eq:dpep}), the assumption $d_{\min}\leq \dpep\leq d_{\max}$  conveys that $\snr^{-d_{\max}}\dotlt P(E)\dotlt \snr^{-d_{\min}}$. By further deploying the assumption $0<d_{\min}$ and some simple manipulations we get $\log P(E)\;\doteq\; -1$, $1-P(E)\;\doteq\; 1$, and $\log(1-P(E))\;\doteq\; 0$. Therefore,
\begin{equation*}
    {\cal H}(P(E))=-P(E)\log(P(E))-(1-P(E))\log(1-P(E))\doteq P(E)\dotlt \snr^{-d_{\min}}.
\end{equation*}
As a result, by noting that $\delta$, $l$, and $R$ are fixed constants we get $\left(\frac{\delta}{R}-\frac{\mathcal{H}(P(E))}{Rl}\right)\doteq 1-P(E)\;\doteq\; 1$. This exponential equality along with (\ref{eq:UB}) establishes the desired result.
\end{proof}

\begin{lemma}[Lower Bound]\label{lemma:4}
For an ISI channel with MMSE SC-FDE receiver we have
\begin{equation*}
    \dout\leq\dpep.
\end{equation*}
\end{lemma}

\begin{proof}
For pairwise error probability analysis, we assess the probability that the transmitted symbol $x(\ell)=s_j$ is erroneously detected as $\tilde x(\ell)=s_k$. By recalling (\ref{eq:MMSE_noise}), the combined channel noise and residual ISI is $\tilde \bn= \sqrt{\snr}(\bW\bH_{\rm eq}-\bI)\bx+\bW \bn$, where it is observed that for any channel realization $H_{\rm eq}$, $\sqrt{\snr}(\bW\bH_{\rm eq}-\bI)$ is deterministic and therefore $\tilde\bn$ inherits all its randomness from $\bn$ and as a result has complex Gaussian distribution. Moreover by using (\ref{eq:MMSE_var}) and following the same approach as in obtaining $\gamma_l^{\rm MMSE}$ in (\ref{eq:MMSE_snr}), the variance of the noise term $\tilde n(\ell)$ is given by
\begin{equation}
    \label{eq:tilde_sigma}\sigma^2_{\tilde\bn}(\ell)=\bbe[|\tilde n(\ell)-\bmu_{\tilde\bn}(\ell)|^2]= \bR_{\tilde\bn}(\ell,\ell)-|\bmu_{\tilde\bn}(\ell)|^2= \frac{1}{L}\sum_{k=1}^L\frac{\snr}{\snr|\lambda_k|^2+1}-|\bmu_{\tilde\bn}(\ell)|^2.
\end{equation}
By noting that $|\bmu_{\tilde\bn}(\ell)|^2$ is the $\ell^{th}$ diagonal element of the matrix $\widehat\bR_{\tilde\bn}$ defined as
\begin{equation}
  \label{eq:MMSE_mean} \widehat\bR_{\tilde\bn}  \dff\bbe[\tilde\bn](\bbe[\tilde\bn])^H  =\snr^{-1}\big[\Heq^H \Heq+\snr^{-1}\bI\big]^{-2},
\end{equation}
and also taking into account that due to the underlying symmetry the diagonal elements of $\widehat\bR_{\tilde\bn}$ are equal we get $|\bmu_{\tilde\bn}(\ell)|^2=\frac{1}{L}{\rm tr}(\widehat\bR_{\tilde\bn})$. By recalling the eigen decomposition of $\Heq$ and matrix trace properties, (\ref{eq:tilde_sigma}) and (\ref{eq:MMSE_mean}) establish that
\begin{align}\label{eq:tilde_sigma2}
    \nonumber \sigma^2_{\tilde\bn}(\ell) & =\frac{1}{L}\sum_{k=1}^L\frac{\snr}{\snr|\lambda_k|^2+1}-\frac{1}{L}{\rm tr}(\widehat\bR_{\tilde\bn})\\
    \nonumber &=\frac{1}{L}\sum_{k=1}^L\frac{\snr}{\snr|\lambda_k|^2+1}- \frac{\snr^{-1}}{L}{\rm tr}\left(\big[\Heq^H \Heq+\snr^{-1}\bI\big]^{-2}\right)\\
    \nonumber &=\frac{1}{L}\sum_{k=1}^L\frac{\snr}{\snr|\lambda_k|^2+1}- \frac{\snr^{-1}}{L}{\rm tr}\left(\big[\bLambda^H\bLambda+\snr^{-1}\bI\big]^{-2}\right)\\
     &=\frac{1}{L}\sum_{k=1}^L\frac{\snr}{\snr|\lambda_k|^2+1}- \frac{1}{L}\sum_{k=1}^L\frac{\snr}{(\snr|\lambda_k|^2+1)^2}= \frac{1}{L}\sum_{k=1}^L\frac{\snr^2|\lambda_k|^2}{(\snr|\lambda_k|^2+1)^2}.
\end{align}
On the other hand, by defining $e_{kj}\dff\frac{s_k-s_j}{|s_k-s_j|}$, the probability of erroneous detection for channel realization $H$ is
\begin{eqnarray*}
  \pr\big[s_j\rightarrow s_k\;|\;\bH=H\big] &=&
  \pr\bigg[\frac{\snr}{4}|s_k-s_j|^2\leq |e^*_{kj}(\tilde y(\ell)-\sqrt{\snr}s_j)|^2\;\Big|\;\bH=H\bigg]\\
  & \leq &
  \pr\bigg[\frac{\snr}{4}|s_k-s_j|^2\leq |\tilde n(\ell)|^2\;\Big|\;\bH=H\bigg],
\end{eqnarray*}
where the inequality holds since $|e^*_{kj}(\tilde y(\ell)-\sqrt{\snr}s_j)\leq |e^*_{kj}||\tilde y(\ell)-\sqrt{\snr}s_j|=|\tilde y(\ell)-\sqrt{\snr}s_j|=|\tilde n(\ell)|$. Now, let us denote the real and imaginary parts of $\tilde n(\ell)$ by $\tilde n_r(\ell)\sim{\cal N}(\mu_r(\ell),\sigma^2_r(\ell))$ and $\tilde n_i(\ell)\sim{\cal N}(\mu_i(\ell),\sigma^2_i(\ell))$, respectively, based on which we have
\begin{equation*}
   \bigg\{\frac{\snr}{4}|s_k-s_j|^2\leq |\tilde n(\ell)|^2\bigg\} \subset \bigg\{\frac{\snr}{16}|s_k-s_j|^2\leq |\tilde n_r(\ell)|^2\bigg\} \bigcup\bigg\{\frac{\snr}{16}|s_k-s_j|^2\leq |\tilde n_i(\ell)|^2\bigg\}.
\end{equation*}
Therefore, by taking into account that $\tilde n_r$ and $\tilde n_i$ have Gaussian distribution and applying the property of the Gaussian tail function $Q(x)\leq\exp(-x^2/2)$ for the pairwise error probability we get
\begin{align}
  \nonumber\pr\big[s_j\rightarrow s_k\;|\;\bH=H\big] & \leq \exp\left(-\frac{(\frac{\sqrt{\snr}}{4}|s_k-s_j|-\mu_r(\ell))^2}{\sigma^2_r(\ell)}\right)+ \exp\left(-\frac{(\frac{\sqrt{\snr}}{4}|s_k-s_j|+\mu_r(\ell))^2}{\sigma^2_r(\ell)}\right)\\
  \nonumber &+ \exp\left(-\frac{(\frac{\sqrt{\snr}}{4}|s_k-s_j|-\mu_i(\ell))^2}{\sigma^2_i(\ell)}\right)+ \exp\left(-\frac{(\frac{\sqrt{\snr}}{4}|s_k-s_j|+\mu_i(\ell))^2}{\sigma^2_i(\ell)}\right)\\
  \nonumber & \leq \exp\left(-\frac{(\frac{\sqrt{\snr}}{4}|s_k-s_j|-\mu_r(\ell))^2}{\sigma^2_{\tilde\bn}(\ell)}\right)+ \exp\left(-\frac{(\frac{\sqrt{\snr}}{4}|s_k-s_j|+\mu_r(\ell))^2}{\sigma^2_{\tilde\bn}(\ell)}\right)\\
  \label{eq:PEP} &+ \exp\left(-\frac{(\frac{\sqrt{\snr}}{4}|s_k-s_j|-\mu_i(\ell))^2}{\sigma^2_{\tilde\bn}(\ell)}\right)+ \exp\left(-\frac{(\frac{\sqrt{\snr}}{4}|s_k-s_j|+\mu_i(\ell))^2}{\sigma^2_{\tilde\bn}(\ell)}\right),
\end{align}
where the las step holds as $\sigma^2_{\tilde\bn}(\ell)=\sigma^2_r(\ell)+\sigma^2_i(\ell)\geq \sigma^2_r(\ell),\sigma^2_i(\ell)$. Now we show that $\mu_r(\ell)\dotlt\snr^{\frac{1}{2}}$ and $\mu_i(\ell)\dotlt\snr^{\frac{1}{2}}$. Recall that, as given in (\ref{eq:MMSE_mean1}), $\bmu_{\tilde\bn}=-\snr^{-\frac{1}{2}}\big[\Heq^H \Heq+\snr^{-1}\bI\big]^{-1}\bx$ and consider the following decomposition
\begin{equation*}
    \big[\Heq^H \Heq+\snr^{-1}\bI\big]^{-1}=\bQ^H\big[\bLambda^H\bLambda+\snr^{-1}\bI\big]^{-1}\bQ=  \bQ^H\bigg[{\rm diag}\Big\{\frac{1}{|\lambda_k|^2+\snr^{-1}}\Big\}\bigg]\bQ.
\end{equation*}
Note that $|\lambda_k|^2+\snr^{-1}\dotgt\snr^{-1}$ or equivalently $\frac{1}{|\lambda_k|^2+\snr^{-1}}\dotlt\snr$. Therefore, all elements of the matrix $\pm \bQ^H\big[\bLambda^H\bLambda+\snr^{-1}\bI\big]^{-1}\bQ\bx$, being linear combinations of $\{\frac{1}{|\lambda_k|^2+\snr^{-1}}\}$, cannot grow with $\snr$ faster than $\snr$, and therefore, the elements of $\pm \snr^{-\frac{1}{2}}\big[\Heq^H \Heq+\snr^{-1}\bI\big]^{-1}\bx$ cannot grow with $\snr$ faster than $\snr^{-\frac{1}{2}}$, i.e., $\pm \bmu_{\tilde\bn}(\ell)\dotlt\snr^{\frac{1}{2}}$ and thereof, $\snr^{\frac{1}{2}}\pm\bmu_{\tilde\bn}(\ell)\doteq\snr^{\frac{1}{2}}$. The same result is concluded for $\mu_r(\ell)$ and $\mu_i(\ell)$, being the real and imaginary parts of $\bmu_{\tilde\bn}(\ell)$.

As a result, for any $s_k$ and $s_j$, $\frac{\sqrt{\snr}}{4}|s_k-s_j|\pm\mu_r(\ell)\;\doteq\; \snr^{\frac{1}{2}}\pm\mu_r(\ell)\;\doteq\;\snr^{\frac{1}{2}}$ and similarly $\frac{\sqrt{\snr}}{4}|s_k-s_j|\pm\mu_i(\ell)\;\doteq\; \snr^{\frac{1}{2}}$. Hence, from (\ref{eq:tilde_sigma2}) and (\ref{eq:PEP}) we get
\begin{eqnarray*}
  \nonumber\pr\big[s_j\rightarrow s_k\;|\;\bH=H\big]  \dotlt 4\exp\left(-\frac{\snr}{\sigma^2_{\tilde\bn}(\ell)}\right)\doteq \exp\bigg[-\bigg(\frac{1}{L}\sum_{k=1}^L\frac{\snr|\lambda_k|^2}{(\snr|\lambda_k|^2+1)^2}\bigg)^{-1}\bigg].
\end{eqnarray*}
By denoting the error event by $E$ and applying the union bound, for data rate $R$ and uncoded transmission ($l=1$) we get
\begin{align}\label{eq:union_bound}
     \pr\Big(E\med \bH=H\Big)&\dotlt 2^{R}\exp\bigg[-\bigg(\frac{1}{L}\sum_{k=1}^L\frac{\snr|\lambda_k|^2}{(\snr|\lambda_k|^2+1)^2}\bigg)^{-1}\bigg].
\end{align}
Next, in order to find the exponential order of $\pe=P(E)$ we first find the probability of occurring an error while there is no outage, i.e., $\pr(E,\bar O\med \bH=H\Big)$ where $\bar O$ denotes the non-outage event which based on (\ref{eq:Pout}) is given by
\begin{equation}\label{eq:O1}
     \bar O=\bigg\{\frac{1}{L} \sum_{k=1}^{L} \frac{1}{1+\snr|\lambda_k|^2}<2^{-R}\bigg\}
     \quad\Rightarrow\quad \exp\bigg[2^{R}-\bigg(\frac{1}{L}\sum_{k=1}^L\frac{1}{\snr|\lambda_k|^2+1}\bigg)^{-1}\bigg]\leq 1.
\end{equation}
By representing the channel matrix with the exponential orders of the
eigenvalues $\{\alpha_k\}$ and recalling the equality-in-the-limit
\begin{equation}\label{eq:limit}
 \frac{1}{1+\snr|\lambda_k|^2} \doteq \begin{cases}
 \snr^{\alpha_k-1} &  \alpha_k<1 \\
 1& \alpha_k>1
 \end{cases}\quad\mbox{for}\;\;k=1,\dots,L,
\end{equation}
and by following the same line of argument as in Lemma \ref{lemma:1} we get
\begin{equation}\label{eq:O2}
     \bar O=\big\{M(\balpha)\leq \lfloor L2^{-R}\rfloor\big\},
\end{equation}
where we had defined $M(\balpha) = \sum_{\alpha_k>1} 1$ in (\ref{eq:M}). Note that for the region $\{\balpha\med M(\balpha)=0\}$ there will be no outage for any rate as for any $R>0$ we have $\lfloor 2^{-R}\rfloor\geq M(\balpha)=0$. On the other hand, in the region $\{\balpha\med M(\balpha)\geq 1\}$ there will be outage for the rates $R\leq \log L$. We investigate these two regions separately.

For the region $\{\balpha\med M(\balpha)=0\}$, over which we have $\max_i\alpha_k<1$, from (\ref{eq:union_bound}) and (\ref{eq:limit}) and some simple manipulations we get
\begin{equation}\label{eq:PEP4}
 \pr(E,\;\bar{O}\med M(\balpha)=0)\;\dotlt \;
 2^R\exp\bigg[-L\bigg(\snr^{\max_k\alpha_k-1}\bigg)^{-1}\bigg]=2^R\exp\bigg[-L\snr^{1-\max_k\alpha_k}\bigg].
\end{equation}
Since the growth of the exponential function is faster than polynomial functions and $1-\max_i\alpha_k>0$, we get
\begin{equation}\label{eq:PEP5}
 \lim_{\snr\rightarrow\infty}\frac{\exp\bigg[-L\snr^{1-\max_k\alpha_k}\bigg]}{\snr^{-(\nu+1)}}=0,
\end{equation}
which in turn provides that
\begin{equation}\label{eq:PEP5}
     \pr(E,\;\bar{O}\med M(\balpha)=0)\;\dotlt \;\exp\bigg[-L\snr^{1-\max_k\alpha_k}\bigg]\dotlt \snr^{-(\nu+1)}.
\end{equation}
Next, we show the same result for the region $\{\balpha\med M(\balpha)\geq 1\}$. We can rewrite (\ref{eq:union_bound}) as
\begin{align}\label{eq:union_bound2}
    \nonumber \pr\Big(E\med \bH=H\Big)&\dotlt 2^{R}\exp\bigg[-\bigg(\frac{1}{L}\sum_{k=1}^L\frac{\snr|\lambda_k|^2}{(\snr|\lambda_k|^2+1)^2}\bigg)^{-1}\bigg]\\
    \nonumber&\leq\exp\bigg[2^{R}-\bigg(\frac{1}{L}\sum_{k=1}^L\frac{1}{\snr|\lambda_k|^2+1}- \frac{1}{L}\sum_{k=1}^L\frac{1}{(\snr|\lambda_k|^2+1)^2}\bigg)^{-1}\bigg]\\
    \nonumber &=\underset{\leq 1\;\textrm{in the non-outage region}\; \bar O\;\textrm{from (\ref{eq:O1}) }}{\underbrace{ \exp\bigg[2^{R}-\bigg(\frac{1}{L}\sum_{k=1}^L\frac{1}{\snr|\lambda_k|^2+1}\bigg)^{-1}\bigg]}}\\ &\qquad\times\exp\bigg[\bigg(\frac{1}{L}\sum_{k=1}^L\frac{1}{\snr|\lambda_k|^2+1}\bigg)^{-1}- \bigg(\frac{1}{L}\sum_{k=1}^L\frac{\snr|\lambda_k|^2}{(\snr|\lambda_k|^2+1)^2}\bigg)^{-1}\bigg].
\end{align}
Therefore, for the region $\{\balpha\med M(\balpha)\geq 1\}$ we get
\begin{align*}
 \pr(E,\;\bar{O}\med M(\balpha)\geq 1)&\;\dotlt \;
 \exp\bigg[\bigg(\frac{1}{L}\sum_{k=1}^L\frac{1}{\snr|\lambda_k|^2+1}\bigg)^{-1}- \bigg(\frac{1}{L}\sum_{k=1}^L\frac{\snr|\lambda_k|^2}{(\snr|\lambda_k|^2+1)^2}\bigg)^{-1}\bigg]\\
 &= \exp\left[-\frac{\bigg(\frac{1}{L}\sum_{k=1}^L\frac{1}{(\snr|\lambda_k|^2+1)^2}\bigg)} {\bigg(\frac{1}{L}\sum_{k=1}^L\frac{1}{\snr|\lambda_k|^2+1}\bigg) \bigg(\frac{1}{L}\sum_{k=1}^L\frac{\snr|\lambda_k|^2}{(\snr|\lambda_k|^2+1)^2}\bigg)}\right]\\
 &\doteq \exp\left[-\frac{LM(\balpha)} {M(\balpha) \snr^{-\min_k|1-\alpha_k|}}\right]\qquad(\mbox{note that }\;M(\balpha)\geq 1)\\
 &\doteq \exp\left[-L \snr^{\min_k|1-\alpha_k|}\right].
\end{align*}
By noting that $|1-\alpha_k|>0$ and following the same line of argument as in (\ref{eq:PEP4})-(\ref{eq:PEP5}) we find that
\begin{equation}\label{eq:PEP6}
     \pr(E,\;\bar{O}\med M(\balpha)\geq 1)\;\dotlt \;\exp\bigg[-L\snr^{1-\max_k\alpha_k}\bigg]\dotlt \snr^{-(\nu+1)}.
\end{equation}
Therefore, if we denote the pdf of $\balpha$ by $p(\balpha)$, and invoke the results of (\ref{eq:PEP5}) and (\ref{eq:PEP6}) we get
\begin{eqnarray*}
 \pr(E,\;\bar{O})&=&\int_{M(\balpha)=0}\pr\big(E,\;\bar{O}\med M(\balpha)=0\big) p(\balpha)\; d\balpha + \int_{M(\balpha)\geq 1}\pr\big(E,\;\bar{O}\med M(\balpha)\geq 1\big) p(\balpha)\; d\balpha\\
 &\;\dotlt\; & \int_{M(\balpha)=0}\snr^{-(\nu+1)}p(\balpha)\; d\balpha + \int_{M(\balpha)\geq 1}\snr^{-(\nu+1)} p(\balpha)\; d\balpha\\
 &=&\snr^{-(\nu+1)}\int_{M(\balpha)=0}p(\balpha)\; d\balpha + \int_{M(\balpha)\geq 1}p(\balpha)\; d\balpha\\
 &=& \snr^{-(\nu+1)}.
\end{eqnarray*}
Finally, by taking into account that we always have $\dout\leq \nu+1$ (based on (\ref{eq:cor1})) we get
\begin{eqnarray}
\nonumber\pe&=& \pr(E\med O)\cdot\pout +\pr(E,\;\bar{O}) \\
\nonumber&\leq& \pout+\pr(E,\;\bar{O})\\
\nonumber&\dotlt&\snr^{-\dout}+\snr^{-(\nu+1)}\\
\nonumber &\doteq&\snr^{-\dout}\\
&=& \pout.
\end{eqnarray}
Therefore, we always have $\dpep\geq\dout$, which concludes the proof of the lemma.
\end{proof}

Lemmas~\ref{lemma:3} and~\ref{lemma:4}, in conjunction with
Corollary~\ref{cor1} characterize the diversity order achieved in ISI
channels with MMSE SC-FDE which is stated in the following theorem.

\begin{theorem}[MMSE Diversity Gain]
For an ISI channel with MMSE SC-FDE, the average pairwise error
probability (PEP) and the outage probability are exponentially equal and the diversity gain is $\dpep=\dout$, where $\dout$ is given in (\ref{eq:cor1}).
\end{theorem}
\begin{proof}
The characterization of $\dout$ given in (\ref{eq:cor1}) provides that $\dout\geq 1$. Therefore, by applying Lemma \ref{lemma:4} we find that $\dpep\geq 1$. On the other hand as the diversity gain cannot exceed the degrees of freedom $(\nu+1)$ we also find that $1\leq\dpep\leq (\nu+1)$. Therefore, by selecting $d_{\min}=1$ and $d_{\max}=\nu+1$ the conditions of Lemma \ref{lemma:3} are satisfied and as a result we find $\dout\geq\dpep$. This result in conjunction with the inequality $\dpep\geq\dout$ from Lemma (\ref{lemma:4}) concludes the desires result.
\end{proof}

\section{Zero-Forcing Diversity}
\label{sec:diversity_ZF}

Zero-forcing (ZF) equalizers invert the channel and remove all ISI
from the received values. For the system defined in~(\ref{eq:model1})
the ZF linear equalizer is
\begin{equation*}
 \bW_{\rm ZF}=\Heq^{-1}=\bQ^H\bLambda^{-1}\bQ,
\end{equation*}
and the equalizer taps are $\lambda_i^{-1}$ as defined n~(\ref{eq:model3}). Thus the equalizer output is
\begin{equation*}
 \tilde {\by}=\sqrt{\snr}\,\bx+\Heq^{-1}\bn,
\end{equation*}
where the noise term $\tilde{\bn}=\Heq^{-1}\bn$ has
covariance matrix
\begin{equation}
 \label{eq:zf_nosie}
 \bR_{\tilde \bn}=\bbe\big[\tilde \bn\tilde \bn^H\big]=
 \bQ^H(\bLambda\bLambda^H)^{-1}\bQ.
\end{equation}
Since all the diagonal elements of the matrix $\bR_{\tilde \bn}$ are equal, the decision-point SINR for detecting symbol $x(\ell), 1\leq \ell\leq L$ is given by
\begin{align*}
 \gamma_\ell^{\rm ZF}&=\frac{\snr}{\frac{1}{L}{\rm tr}[\bR_{\tilde \bn}]}
 =\frac{\snr}{\frac{1}{L}{\rm tr}[\bQ(\bLambda\bLambda^H)^{-1} \bQ^{\rm
 H}]}=\frac{\snr}{\frac{1}{L}{\rm tr}[(\bLambda\bLambda^H)^{-1}]}
 =\bigg[\frac{1}{L}\sum_{k=1}^L\frac{1}{\snr|\lambda_k|^2}\bigg]^{-1}.
\end{align*}

For ZF SC-FDE receiver, the effective mutual information between $\bx$ and $\tilde\by$ is equal to the sum of the mutual information of their components given by
\begin{equation}
 \label{eq:zf_mutual}
 \info_{\rm ZF}(\bx;\boldsymbol{\tilde y})
=\log\bigg[1+\frac{1}{\frac{1}{L}\sum_{k=1}^L\frac{1}{\snr|\lambda_k|^2}}\bigg].
\end{equation}

\begin{theorem}
\label{th:3}

In an ISI channel with memory length $\nu$, transmission block length $L$, data rate $R$, and ZFE SC-FDE receiver, the diversity gain is always 1, i.e., $\pe\doteq \snr^{-1}$.
\end{theorem}
\begin{proof}
Given the mutual information for ZF equalization in
(\ref{eq:zf_mutual}) the outage probability is
\begin{equation}
 \label{eq:th3_proof1}
 \pout=\pr\bigg[\sum_{k=1}^L\frac{1}
 {\snr|\lambda_k|^2}>\frac{L}{2^R-1}\bigg].
\end{equation}
By using the definition of $\alpha_i$ and replacing $|\lambda_i|^2\doteq\snr^{-\alpha_i}$ we get
\begin{eqnarray}
 \nonumber\pout&=&
 \pr\bigg[\sum_{k=1}^L\frac{1}{\snr|\lambda_k|^2}> \frac{L}{2^R-1}\bigg]\\
 \nonumber &\doteq&\pr\bigg[\sum_{k=1}^L
 \snr^{(\alpha_k-1)}>\frac{L}{2^R-1}\bigg]\\
 \nonumber &=&\pr[\max_k \{\alpha_k-1\}>0]\\
 \nonumber &=&\pr[\max_k \{\alpha_k\}>1]\\
 \label{eq:th3_proof2}&\geq&\pr[\alpha_1>1]\\
 \label{eq:th3_proof3}&\doteq& \snr^{-1},
\end{eqnarray}
where (\ref{eq:th3_proof2}) is obtained by noting that the event $\{\alpha_1>1\}$ is a subset of the event $\{\max_k\alpha_k>1\}$ which provides that $P(\max_k\alpha_k>1)\geq P(\alpha_1>1)\doteq\snr^{-1}$.
Therefore unlike MMSE equalization, for ZF equalization $\dout$ cannot
exceed 1. This result holds for all rates, block transmission lengths
and is independent of channel memory length. The result of
Lemma~\ref{lemma:3} holds for ZF SC-FDE too, concluding that
$\pe\dotgt\pout\dotgt\snr^{-1}$. It is easy to verify that
diversity gain 1 is always achievable, which concludes the proof.
\end{proof}

\section{Simulation Results}
\label{sec:simulations}
\begin{figure}[t]
\centering
\includegraphics[width=5in]{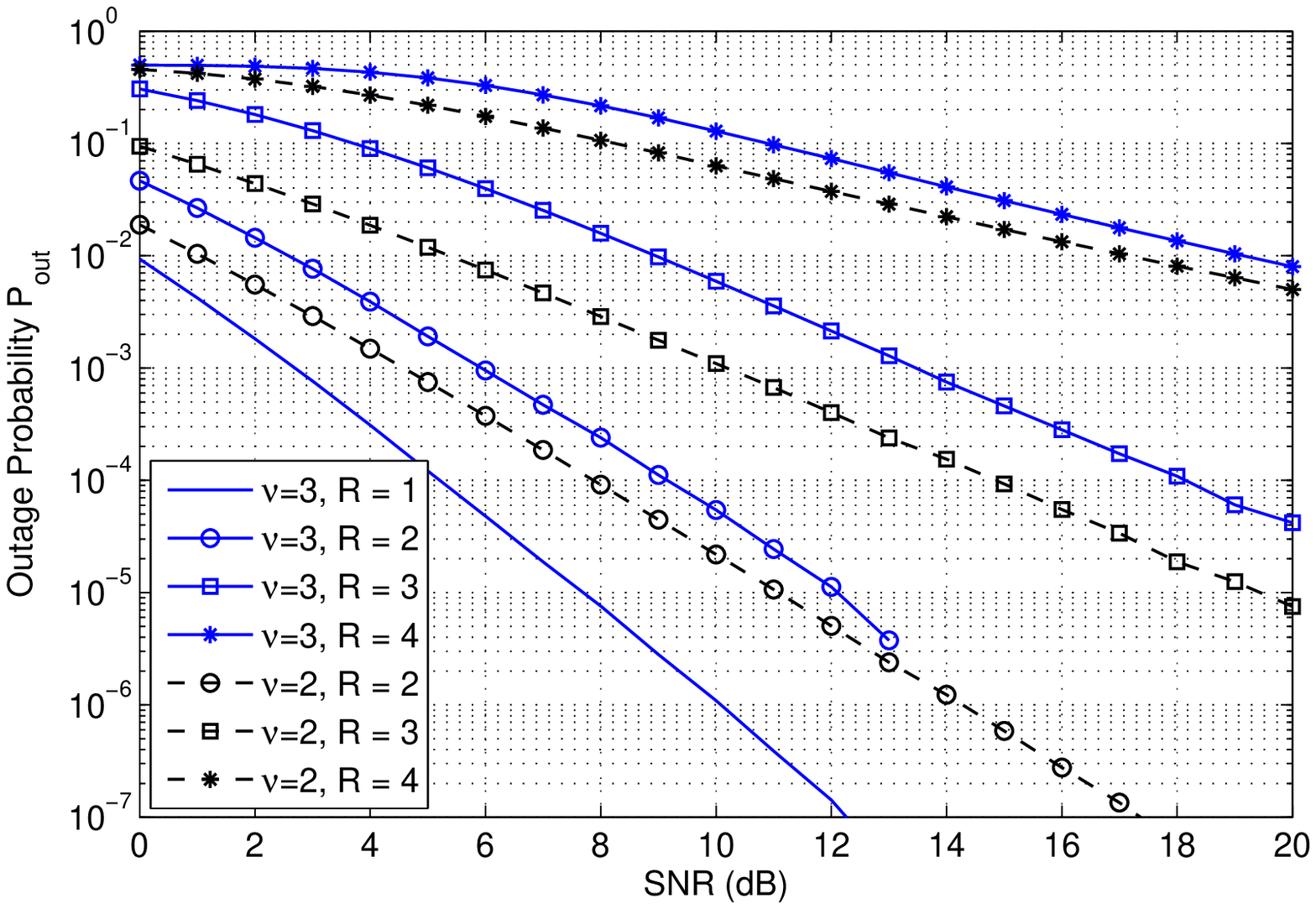}\\
\caption{Outage probability for MMSE SC-FDE block transmission in a channels with memory lengths $\nu=2, 3$, block length $L=10$ and different data rates $R=1,2,3,4$.}\label{fig:Pout}
\end{figure}

In this section we provide numerical evaluation and simulations results for assessing the outage and pairwise error probabilities. Figure \ref{fig:Pout} depicts the numerical evaluation of the outage probability for MMSE receivers given in (\ref{eq:Pout}). We consider block transmissions of length $L=10$ for frequency selective channels with memory lengths $\nu=2,3$. Based on the numerical evaluations we find that for $\nu=2$ and rates $R=2,\;3,\;4$, the negative of the exponential order fo outage probabilities are $d=3,\;2,\;1$, respectively. Note that for for $\nu=2$ and $L=10$, the rate intervals characterized in (\ref{eq:cor1}) for achieving diversity gains 3, 2, 1 are (0, 2.32], (2.32, 3.32], and (3.32, $\infty$), respectively, which anticipate achieving the same diversity gains as achieved by the numerical evaluations. The same evaluations is carried out for the case of $\nu=3$ and $L=4$ as well where it is observed that for $R=1,\;2,\;3,\;4$ the diversity gains are $d=4,\; 3,\;2,\;1$, respectively and match the results expected from (\ref{eq:cor1}) from which we obtain the rate intervals (0, 1.73], (1.73, 2.32], (2.32, 3.32], and (3.32, $\infty$).

\begin{figure}[t]
\centering
\includegraphics[width=5in]{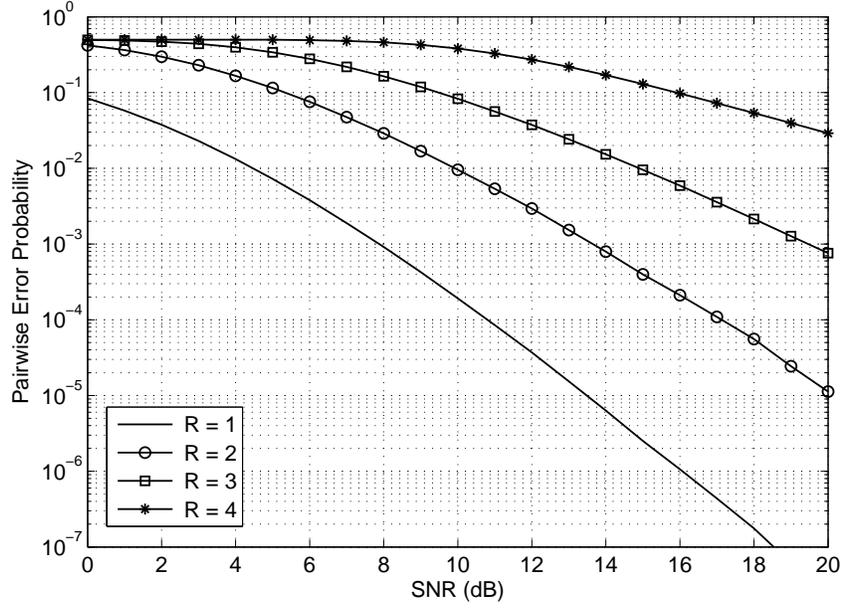}\\
\caption{Achievable diversity order in MMSE SC-FDE block transmission in channels with memory length $\nu=3$, block length $L=10$ and different data rates $R=1,2,3,4$.}\label{fig:PEP}
\end{figure}

For examining the asymptotic equivalent of outage and pairwise error probabilities, Fig. \ref{fig:PEP} illustrates the simulation results on the pairwise error probability. We have considered the setting $\nu=3$ and $L=$ and uncoded transmission where the symbols are drawn from $2^{R}$-PSK constellations for $R=1,\dots,4$. It is observed that the achievable diversity gain for the rates $R=$ 1, 2, 3, 4, are $d=$ 4, 3, 2, 1, respectively.

\begin{figure}[t]
\centering
\includegraphics[width=5in]{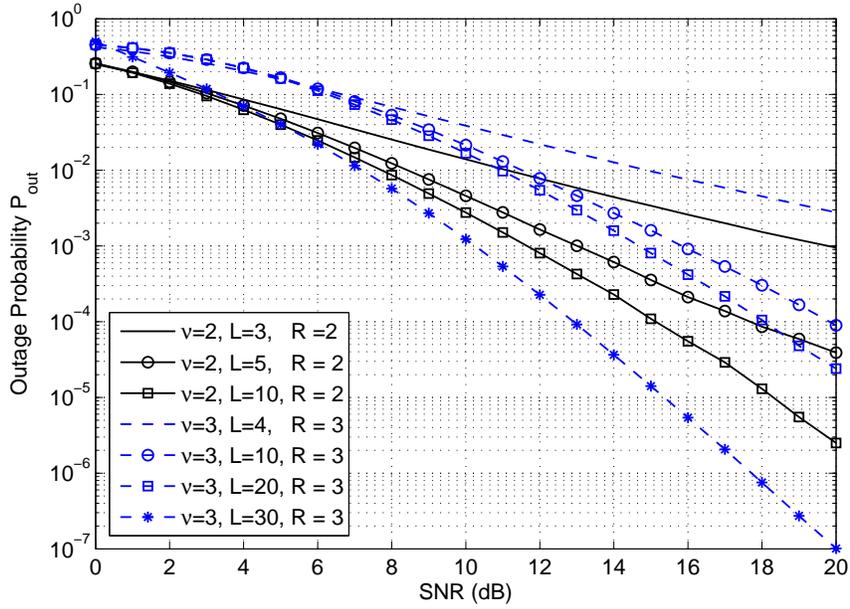}
\caption{The effect of transmission block length on the diversity order for the settings $(\nu,R)=(2,2)$ and $(\nu,R)=(3,3)$.}
\label{fig:length}
\end{figure}

In Fig. \ref{fig:length} we provide the numerical evaluations of the outage probability for showing the effect of varying transmission block lengths. It is demonstrated that for fixed data rates, it is possible to span the entire range of diversity gains by controlling the transmission block lengths. The evaluations are provided for the settings $(\nu,R)=(2,2)$ and $(\nu,R)=(3,3)$.

\begin{figure}[t]
\centering
\includegraphics[width=5in]{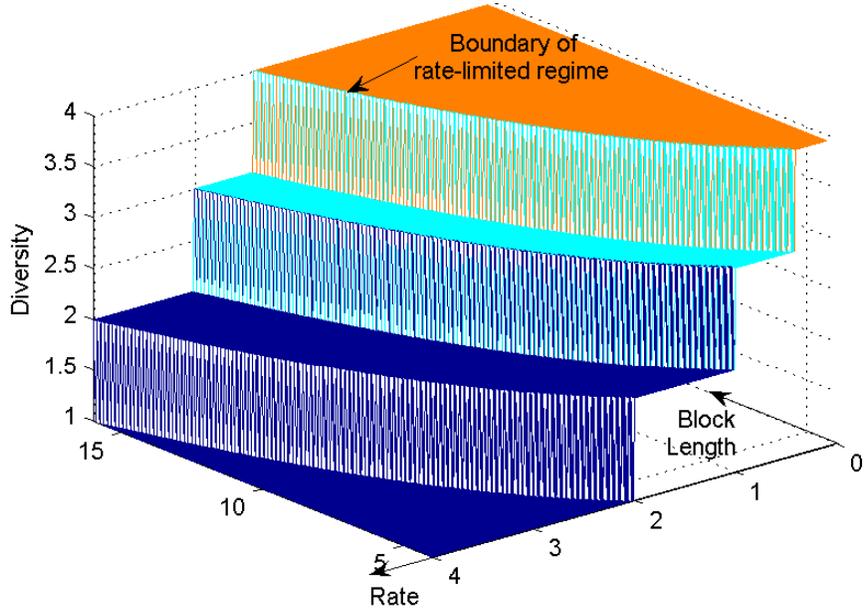}
\caption{The tradeoff between diversity, rate, and block length for MMSE SC-FDE.}
\label{fig:3D}
\end{figure}

\begin{figure}[t]
\centering
\includegraphics[width=5in]{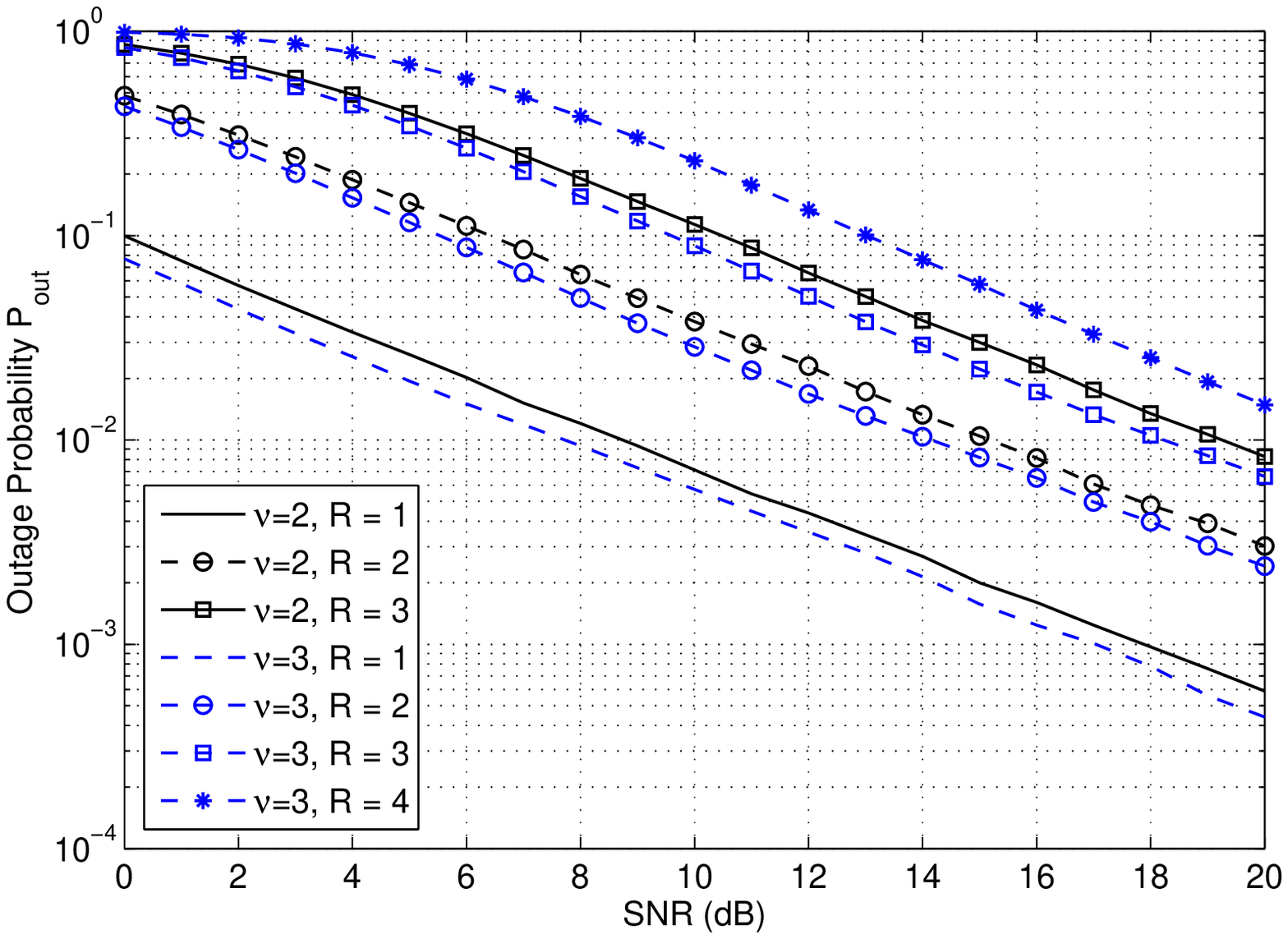}
\caption{Average error probability for ZF SC-FDE block transmission for channel memory lengths $\nu=2,3$, block length $L=10$, and data rates $R=1,2,3,4$.}
\label{fig:zf}
\end{figure}

The tradeoff between diversity order, data rate, channel memory length, and transmission block length is demonstrated in Fig.~\ref{fig:3D} for a representative example and finally Fig.~\ref{fig:zf} shows the simulation results on the diversity order achieved by ZF SC-FDE receivers. It is shown that the diversity order for different channel memory lengths, data rates, and transmission block lengths $L=10$ is 1.

\section{Discussion and Conclusion}
\label{sec:conclusion}
In this paper we analyze the diversity of single-carrier cyclic-prefix block transmission with frequency-domain linear equalization. We show that MMSE SC-FDE may not fully capture the inherent frequency diversity of the ISI channels, depending on the system settings. We show that for such receivers, there exist a tradeoff between achievable diversity order, data rate and transmission block length. At high rates and low block-lengths, only diversity 1 is achieved, but by increasing the transmission block length and/or decreasing data rate, diversity order can be increased up to a maximum level of $\nu+1$, where $\nu$ is the channel memory length. We characterize the dependence on these two parameters in our results. Specifically, it is demonstrated that for MMSE SC-FDE, the results admit an interpretation in terms of operating regimes. As long as $R\leq\log\frac{L}{\nu}$, full diversity is achieved regardless of the exact value of the rate. When $R>\log\frac{L}{\nu}$ we are in a \emph{rate-limited} regime where the diversity is affected by rate. In this regime, to maintain a given diversity while increasing the rate, each additional bit of spectral efficiency must be offset by at most doubling the block length. Naturally the block length cannot exceed the coherence time of the channel, thus putting practical limits on the performance of the equalizer.

We also prove that for zero-forcing SC-FDE, the diversity order is always one, independently of channel memory, transmission block length, or data rate.

For clarity and ease of exposition, the rates $R$ in this paper do not
include the fractional rate loss incurred by the cyclic prefix. Once the
fractional rate loss is included, the overall throughput will be equal
to $R'= \frac{L}{L+\nu} R$ which can be easily factored into all
results.

Finally we would like to remark that the shorter version of this paper \cite{ali:gc07}, which provides the outage analysis for MMSE equalizers, differs with the current paper in the following directions. First, \cite{ali:gc07} only treats MMSE equalizers whereas in this paper we have treated both MMSE and ZF equalizers. Secondly and more importantly, the analysis in \cite{ali:gc07} characterizes only the outage probability and its asymptotic behavior which does not suffice to obtain the diversity gain and, as discussed in Section \ref{sec:PEP}, requires further analysis to establish the connection between the outage probability and the pair-wise error probability. Finally, we have provided a new and more intuitive proof for lemmas \ref{lemma:1} and \ref{lemma:2}, which have key roles in characterizing the outage probability.

\appendix

\section{Proof of Lemma \ref{lemma:2}}
\label{app:lemma3}

We start by showing that for any integer multiplier of $L$ denoted by $\tilde L=TL$, where $T\in\mathbb{N}$, and for any real-valued $m\in(0,\nu+1)$ we have
\begin{equation}
 \label{eq:lemma3_proof1}
 \pr\bigg[\sum_{k=1}^{\tilde L}\frac{1}{1+\snr|\tilde\lambda_k|^2}>m\bigg]
 \doteq
 \pr\bigg[\sum_{k=1}^{L}\frac{1}{1+\snr|\lambda_k|^2}>m\bigg],
\end{equation}
where we have defined
\begin{equation*}
    \tilde\bg_{1\times\tilde L}\dff[h_0,\dots,h_{\nu},\underset{\tilde L-\nu-1}{\underbrace{0,\dots,0}}]\quad\mbox{and}\quad \{\tilde\lambda_i\}_{1\times\tilde L}\dff{\rm DFT}(\tilde\bg),
\end{equation*}
and therefore, $\tilde g_{1\times\tilde L}$ is a zero-padded version of $g_{1\times L}$. Note that zero padding and applying a larger DFT size ($\tilde L$) is equivalent to sampling the Fourier transform of the $L$ data points at $\tilde L$ points. Based on the given set of DFT points $\{\lambda_k\}$ we can characterize the Fourier transform of $\bg$ denoted by $G(\omega)$ at any specific frequency $\omega$ via
\begin{equation}\label{eq:lemma3_proof3}
    G(\omega)=\frac{1}{L}\sum_{i=1}^{L} \lambda_i\frac{1-e^{-jL\omega}}{1-e^{- j(\omega-\frac{2\pi (i-1)}{L})}}.
\end{equation}
 Therefore the DFT points $\{\tilde\lambda_k\}$ can be found by sampling the Fourier Transform $G(\omega)$ at frequenies $\omega=2\pi \frac{k-1}{\tilde L}$ for $k=1,\dots,\tilde L$. Therefore, we can describe the DFT points $\{\tilde\lambda_k\}$ in terms of $\{\lambda_k\}$ as
\begin{equation}\label{eq:lemma3_proof4}
    \tilde\lambda_k=\sum_{i=1}^{L} \lambda_i\;\underset{\dff \;\gamma_i}{\underbrace{\frac{1}{L}\cdot\frac{1-e^{-j \frac{(k-1)2\pi L}{\tilde L}}}{1-e^{- j(\frac{2\pi (k-1)}{\tilde L}-\frac{2\pi (i-1)}{L})}}}},\quad\mbox{for}\;k=1,\dots,\tilde L.
\end{equation}
Moreover, since $\tilde L=TL$ we have
\begin{equation}\label{eq:lemma3_proof5}
    \tilde\lambda_{T(k-1)+1}=\lambda_{k},\quad\mbox{for}\;k=1,\dots,L.
\end{equation}
By defining $\alpha_k\dff-\frac{\log|\lambda_k|^2}{\log\snr}$ and $\tilde\alpha_k\dff-\frac{\log|\tilde\lambda_k|^2}{\log\snr}$, for $k=1,\dots, L$ from (\ref{eq:lemma3_proof5}) we get
\begin{equation}\label{eq:lemma3_proof6}
    \tilde\alpha_{T(k-1)+1}=\alpha_{k},\quad\mbox{for}\;k=1,\dots,L.
\end{equation}
Also, from (\ref{eq:lemma3_proof4}) we get
\begin{equation}\label{eq:lemma3_proof7}
    |\tilde\lambda_k|^2=\sum_{i=1}^{L} |\gamma_i|^2|\lambda_i|^2+\sum_{i=1}^{L}\sum_{l=1}^{L} \gamma_i\gamma^*_l\lambda_i\lambda^*_l,\quad\mbox{for}\;k=1,\dots,\tilde L.
\end{equation}
Since for any specific $\tilde L$ the coefficients $\{\gamma_k\}$ are constant values, we get
\begin{equation*}
    \lim_{\snr\rightarrow\infty}\frac{\log|\gamma_i|^2|\lambda_i|^2}{\log\snr}= \lim_{\snr\rightarrow\infty}\frac{\log|\gamma_i|^2+\log|\lambda_i|^2}{\log\snr}= \lim_{\snr\rightarrow\infty}\frac{\log|\lambda_i|^2}{\log\snr}\quad\Rightarrow\quad |\gamma_i|^2|\lambda_i|^2\doteq|\lambda_i|^2.
\end{equation*}
Let us also define $A\dff\sum_{i=1}^{L}\sum_{l=1}^{L}\gamma_i\gamma^*_l\lambda_i\lambda^*_l$ and $\alpha_A\dff-\frac{\log |A|}{\log\rho}$ which provides that $|A|=\rho^{-\alpha_A}$. Therefore (\ref{eq:lemma3_proof7}) can be rewritten as
\begin{equation}\label{eq:lemma3_proof8}
    \snr^{-\tilde\alpha_k}\doteq\sum_{i=1}^{L} \snr^{-\alpha_i}+\frac{A}{|A|}\snr^{-\alpha_A}\doteq\snr^{-\min_i\alpha_i}+\frac{A}{|A|}\snr^{-\alpha_A} ,\quad\mbox{for}\;k=1,\dots,\tilde L.
\end{equation}
Note that if $A<0$ we should have $\alpha_A\leq\min_i\alpha_i$ as otherwise for large values of $\snr$ the RHS of (\ref{eq:lemma3_proof8}) will be negative while the LHS is positive. Therefore, for $A<0$ we have $\snr^{-\min_i\alpha_i}+\frac{A}{|A|}\snr^{-\alpha_A}\doteq\snr^{-\min_i\alpha_i}$. On the other hand, for $A\geq 0$ we have $\snr^{-\min_i\alpha_i}+\snr^{-\alpha_A}\dotgt\snr^{-\min_i\alpha_i}$. Hence, in summary we always have
\begin{equation}\label{eq:lemma3_proof9}
    \snr^{-\tilde\alpha_k}\doteq \snr^{-\min_i\alpha_i}+\frac{A}{|A|}\snr^{-\alpha_A}\dotgt\snr^{-\min_i\alpha_i}\quad\Rightarrow\quad\tilde\alpha_k\leq\min_i\alpha_i.
\end{equation}
Now by using (\ref{eq:lemma3_proof6}) and (\ref{eq:lemma3_proof7}) we group the indices of the DFT points into two disjoint sets denoted by ${\cal A}\dff\{T(i-1)+1\med i=1,\dots,L\}$ and ${\cal B}\dff\{1,\dots,\tilde L\}\backslash\{T(i-1)+1\med i=1,\dots,L\}$. Therefore, by taking into account (\ref{eq:lemma3_proof6}) we get
\begin{align}
 \label{eq:lemma3_proof10}
 \nonumber \pr\bigg[\sum_{k=1}^{\tilde L}&\frac{1}{1+\snr|\tilde\lambda_k|^2}>m\bigg]
 = \pr\bigg[\sum_{k\in{\cal A}}\frac{1}{1+\snr|\tilde\lambda_k|^2}+ \sum_{k\in{\cal B}}\frac{1}{1+\snr|\tilde\lambda_k|^2}>m\bigg]\\
 \nonumber &\;\;\doteq
 \pr\bigg[\sum_{k=1}^{L}\frac{1}{1+\snr^{1-\alpha_k}}+\sum_{k\in{\cal B}}\frac{1}{1+\snr^{1-\tilde\alpha_k}}>m\bigg]\\
 \nonumber &=
 \pr\bigg[\sum_{k=1}^{L}\frac{1}{1+\snr^{1-\alpha_k}}+\sum_{k\in{\cal B}}\frac{1}{1+\snr^{1-\tilde\alpha_k}}>m\;\Big|\; \min_i\alpha_i<1\bigg]P(\min_i\alpha_i<1)\\
  &+
 \pr\bigg[\sum_{k=1}^{L}\frac{1}{1+\snr^{1-\alpha_k}}+\sum_{k\in{\cal B}}\frac{1}{1+\snr^{1-\tilde\alpha_k}}>m\;\Big|\; \min_i\alpha_i>1\bigg]P(\min_i\alpha_i>1).
\end{align}
Next, we further simplify the summands in (\ref{eq:lemma3_proof10}). By taking into account that $\tilde\alpha_k\leq\min_i\alpha_i$, conditioning on the event $\{\min_i\alpha_i<1\}$ provides that $\sum_{k\in{\cal B}}\frac{1}{1+\snr^{1-\tilde\alpha_k}}=0$ and the first summand becomes
\begin{align}\label{eq:lemma3_proof11}
    \nonumber \pr\bigg[\sum_{k=1}^{L}\frac{1}{1+\snr^{1-\alpha_k}}+\sum_{k\in{\cal B}}\frac{1}{1+\snr^{1-\tilde\alpha_k}}>m&\;\Big|\; \min_i\alpha_i<1\bigg]\\&=\pr\bigg[\sum_{k=1}^{L}\frac{1}{1+\snr^{1-\alpha_k}}>m\;\Big|\; \min_i\alpha_i<1\bigg].
\end{align}
On the other hand, conditioning on the event $\{\min_i\alpha_i>1\}$ provides that $\sum_{k=1}^{L}\frac{1}{1+\snr^{1-\alpha_k}}=L$ and $\sum_{k=1}^{L}\frac{1}{1+\snr^{1-\alpha_k}}+\sum_{k\in{\cal B}}\frac{1}{1+\snr^{1-\min_i\alpha_i}}\geq L$. Therefore, since $L> m\in(0,\nu+1)$ the second summand becomes
\begin{align}\label{eq:lemma3_proof12}
    \nonumber \pr\bigg[\sum_{k=1}^{L}\frac{1}{1+\snr^{1-\alpha_k}}+\sum_{k\in{\cal B}}\frac{1}{1+\snr^{1-\tilde\alpha_k}}>m & \;\Big|\; \min_i\alpha_i>1\bigg]\\
    & =\pr\bigg[\sum_{k=1}^{L}\frac{1}{1+\snr^{1-\alpha_k}}>m\Big| \min_i\alpha_i>1\bigg]=1.
\end{align}
Combining (\ref{eq:lemma3_proof10})-(\ref{eq:lemma3_proof12}) establishes that
\begin{align}
 \label{eq:lemma3_proof13}
 \nonumber \pr\bigg[\sum_{k=1}^{\tilde L}\frac{1}{1+\snr|\tilde\lambda_k|^2}>m\bigg]
 &= \pr\bigg[\sum_{k=1}^{L}\frac{1}{1+\snr^{1-\alpha_k}}>m\;\Big|\; \min_i\alpha_i<1\bigg]P(\min_i\alpha_i<1)\\
 \nonumber  &+
 \pr\bigg[\sum_{k=1}^{L}\frac{1}{1+\snr^{1-\alpha_k}}>m\;\Big|\; \min_i\alpha_i>1\bigg]P(\min_i\alpha_i>1)\\
 &= \pr\bigg[\sum_{k=1}^{L}\frac{1}{1+\snr|\lambda_k|^2}>m\bigg]
\end{align}
Therefore, to this end we have established that if $L\big|\tilde L$ then for any real-valued $m\in(0,\nu+1)$ we have
\begin{equation*}
 \pr\bigg[\sum_{k=1}^{\tilde L}\frac{1}{1+\snr|\tilde\lambda_k|^2}>m\bigg]
 \doteq
 \pr\bigg[\sum_{k=1}^{L}\frac{1}{1+\snr|\lambda_k|^2}>m\bigg].
\end{equation*}
Now, lets set $\tilde L=L\times L'$. As $L\big|\tilde L$ we have
\begin{equation}
 \label{eq:lemma3_proof14}
 \pr\bigg[\sum_{k=1}^{\tilde L}\frac{1}{1+\snr|\tilde\lambda_k|^2}>m\bigg]
 \doteq
 \pr\bigg[\sum_{k=1}^{L}\frac{1}{1+\snr|\lambda_k|^2}>m\bigg],
\end{equation}
and since $L'\big|\tilde L$ we have
\begin{equation}
 \label{eq:lemma3_proof15}
 \pr\bigg[\sum_{k=1}^{\tilde L}\frac{1}{1+\snr|\tilde\lambda_k|^2}>m\bigg]
 \doteq
 \pr\bigg[\sum_{k=1}^{L'}\frac{1}{1+\snr|\lambda'_k|^2}>m\bigg].
\end{equation}
(\ref{eq:lemma3_proof14}) and (\ref{eq:lemma3_proof15}) together establish the desired result.

\bibliographystyle{IEEEtran}
\bibliography{IEEEabrv,SCFDE-TWC}

\end{document}